\documentclass[prd,twocolumn,showpacs,nofootinbib,floatfix]{revtex4}
\usepackage{graphicx,amsmath,amssymb}
\newcommand{\haak}[1]{\!\left(#1\right)}
\newcommand{\rhaak}[1]{\!\left [#1\right]}
\newcommand{\lhaak}[1]{\left | #1\right |}

\newcommand{\ahaakl}[1]{\left\{#1\right.}
\newcommand{\ahaakr}[1]{\left.#1\right\}}

\newcommand{\half}{\frac{1}{2}}

\newcommand{\erf}{\operatorname{erf}} 
\newcommand{\erfi}{\operatorname{erfi}}
\renewcommand{\Re}{\operatorname{Re}}
\renewcommand{\Im}{\operatorname{Im}}

\begin{document}
\title{Detecting dark matter in electromagnetic field penetration experiments} 
\author{Saibal Mitra}

\affiliation{Instituut voor Theoretische Fysica,\\ Universiteit van Amsterdam,\\
Valckenierstraat 65,\\ 1018 XE Amsterdam,\\ The Netherlands}
\pacs{95.35.+d}
\date{\today}
\begin{abstract}
Dark matter in the form of particles from a hidden mirror sector has recently been proposed as an explanation for the DAMA annual modulation signal. Here one assumes that there exists a small mixing between photons and mirror photons. We show that dark matter with this property can also be detected in electromagnetic field penetration experiments. Such experiments can be used to measure the speed and direction of the dark matter halo wind, the local density, the temperature, and the strength of the photon-mirror photon mixing interaction. An additional result would be a significant improvement of the upper limit on the photon mass.
\end{abstract}
\maketitle

\section{Introduction}
Dark matter may couple to the electromagnetic field. Although charged dark matter is strongly constrained  \cite{chrg}, particles with magnetic and/or electric dipole moments \cite{dip}, axions \cite{ax} and millicharged dark matter \cite{milex} are viable dark matter candidates.
Another dark matter candidate that can couple to the electromagnetic field is mirror matter. Mirror matter is predicted by extensions of the Standard Model that restore parity symmetry \cite{intr,flv}. In these models the Standard Model is appended by a hidden sector which consists of a parity transformed copy of the Standard Model. Parity can either be spontaneously broken or remain unbroken depending on the Higgs potential \cite{brok}. In this article we shall consider the unbroken case where each Standard Model particle has a mirror partner of exactly the same mass. The doubling of the number of relativistic degrees of freedom during Big Bang Nucleosynthesis (BBN) requires that the temperature in the mirror sector be lower than in the ordinary sector by at least a factor of 2. It has been shown that this can be a typical situation after the inflationary epoch \cite{ber}. Various aspects of mirror dark matter are discussed in recent reviews \cite{review,revber} and in Refs.\ \cite{ber,dark}. See Ref.\ \cite{hisrev} for a historical review. Non-minimal mirror matter models consisting of multiple copies of the Standard Model are discussed in Ref.\ \cite{nonmin}.

Mirror matter would be expected to form compact objects. Such objects could be responsible for the large number of observed MACHOs with masses $M\sim 0.5M_{\odot}$ \cite{mstr}, close-in extrasolar planets \cite{planet} and isolated planetary mass objects \cite{isol}. Non-gravitational interactions between mirror particles and ordinary particles could provide for additional signatures of the mirror world. Such interactions are strongly constrained by renormalizability, gauge invariance and parity symmetry. In Refs.\ \cite{flv,flv2} all such interactions were identified. In this paper we will focus on one of the possible interactions that satisfies these constraints, the photon-mirror photon kinetic mixing interaction \cite{bob,glas,eps,flv,flv2}:
\begin{equation}\label{epsln}
\mathcal{L} = \frac{\epsilon}{2}F\cdot F',
\end{equation}
where $F$ ($F'$) is the field strength tensor
for electromagnetism (mirror electromagnetism).
This term causes a mirror electric charge of $q$ appear to be an ordinary electric charge of $\epsilon q$ \cite{flv,bob,glas}. The photon-mirror photon kinetic mixing interaction has been invoked to explain the DAMA annual modulation signal \cite{dama}, gamma ray bursts \cite{511}, problems with the central engine of supernova explosions and the 511 KeV gamma-rays from the galactic bulge \cite{511}, the cosmic coincidence problem \cite{coinc}, the pioneer spacecraft anomaly \cite{pion,eros}, the low number of small craters on the asteroid Eros \cite{eros} and anomalous impact events \cite{eros,anom}. The possible role of neutrino-mirror neutrino oscillations in gamma ray bursts is discussed in Refs.\ \cite{gamm} and neutron-mirror neutron oscillations has been postulated as an explanation for ultra high energy cosmic rays \cite{uhecr}. 

The photon-mirror photon kinetic mixing interaction \eqref{epsln} can affect BBN. The agreement of standard BBN theory with observations implies a limit of $\epsilon\lesssim 3\times 10^{-8}$ \cite{nucleo}. Another important effect of the kinetic mixing term is that it induces orthopositronium-mirror orthopositronium oscillations \cite{glas} which can be detected in vacuum experiments \cite{fg2}.  The oscillations lead to a higher effective decay rate compared to the QED prediction, because the mirror photons from the decays of mirror orthopositronium are not detected. Because collisions between orthopositronium and gas molecules will inhibit the oscillations, it is possible to distinguish this effect from possible invisible decay modes of orthopositronium. In a vacuum cavity experiment performed in 1990 \cite{ortho1} a faster decay rate than the QED prediction was seen. This was also in conflict with the results other experiments in which the cavity was not vacuum and which were consistent with the QED prediction. A value for $\lhaak{\epsilon}$ of about $10^{-6}$ is needed to explain this anomaly \cite{fg1}. In 2003 a new vacuum experiment was carried out \cite{ortho2}, and this time the anomaly was not seen. This latest experiment implies that $\epsilon\lesssim 5\times 10^{-7}$ \cite{review}. A more sensitive experiment is currently being planned \cite{fg2}.

If $\lhaak{\epsilon}\gtrsim 10^{-10}$ then macroscopic mirror fragments can remain close to the Earth's surface and be detected using centrifuges \cite{centr}. It would also be possible to locate such fragments using the fact that these fragments draw in heat from their surroundings, radiating it away as mirror photons and thereby causing an observable cooling effect \cite{cool}\footnote{Since no mirror stars can be inside our solar system, we don't have to consider absorption of mirror photons by the mirror fragment.}. One would also expect observable effects from meteorites \cite{anom,eros} and micrometeorites \cite{micro}. If $\lhaak{\epsilon}\gtrsim 10^{-8}$ then heavy atoms can bind to heavy mirror atoms \cite{bound}. Such bound states could be detected as anomalously heavy isotopes. Other possible effects of bound states were considered in Ref.\ \cite{bob2}.

Another detectable feature of the mirror world is provided by mirror ions in the galactic halo.
As explained in Refs.\ \cite{review,dama}, most of the mirror dark matter in the galactic halo should be in the form of a plasma. It is expected that this plasma consists mainly of mirror helium and heavier elements, because nucleosynthesis in the mirror sector is expected to yield a large abundance of mirror helium \cite{ber,review}. Assuming that the dark matter halo of our galaxy consists of a mirror helium plasma, hydrostatic equilibrium then implies that the temperature of this plasma is about 330 eV \cite{review}. Such a hot plasma will cool itself down rapidly and collapse into a disk, unless a heat source of about $10^{36}-10^{37}$ Watt exists to keep the mirror plasma hot. Supernovae of mirror stars at a rate of about 1 per year could provide for such a heat source \cite{review}. Such a high rate of mirror supernovae could be the result of the expected high mirror helium abundance of mirror stars \cite{star}.

The galactic mirror matter plasma may already have been detected in the DAMA/NaI experiment \cite{dam,dama}. The DAMA/NaI experiment is one of about 20 experiments in which one tries to detect recoils of nuclei caused by collisions with dark matter particles. The positive results of the DAMA/NaI experiment are difficult to reconcile with the negative results of other experiments, such as e.g.\ CDMS II \cite{cdms} if one assumes that dark matter consists of neutralinos.
The crucial difference between neutralinos and mirror ions is that neutralinos are expected to be much heavier. Unlike other detectors, the DAMA detector is capable of detecting recoils caused by mirror ions because it contains the light element sodium. The mirror world explanation for the DAMA results assumes that the detected recoils are caused by mirror oxygen ions or heavier ions colliding with sodium ions in the DAMA detector. Assuming that mirror oxygen is responsible for the DAMA signal, the value for $\lhaak{\epsilon}$ needs to be about $5\times 10^{-9}\sqrt{0.1/\xi}$, where $\xi$ is the relative abundance by mass of mirror oxygen \cite{dama}. Other explanations for the DAMA results include light WIMPs \cite{gelm}, WIMPs with spin dependent interactions \cite{sav}, inelastic dark matter \cite{weiner}, fourth generation neutrinos \cite{khlp}, self-interacting dark matter \cite{mitra}, and pseudoscalar and scalar dark matter \cite{scalar}.

In this article we will explore another way of detecting the mirror plasma. The photon-mirror photon kinetic mixing term \eqref{epsln} couples the mirror plasma to ordinary electromagnetic fields. This means that in very strong electric fields the plasma will become slightly polarized. Mirror charges will be induced which one can attempt to detect inside superconducting cavities.
Strong magnetic fields should also give rise to polarization effects, because the plasma is expected to have a non-zero velocity, $v_{\text{h}}$, w.r.t.\ the Earth. By comparing the results of experiments using magnetic and electric fields one can thus measure both the magnitude and the direction of $v_{\text{h}}$. The velocity of the plasma also makes the polarization effect in purely electric fields slightly anisotropic. We will show that this effect can be used to determine $\lhaak{\epsilon}$, the temperature and the density of the mirror plasma.

\section{Polarizing the mirror plasma}\label{secpol}
Unlike direct detection experiments which are sensitive to the mirror ions, polarization effects are mainly caused by the response of the mirror electrons to the electromagnetic fields. The most important quantities determining this effect are the mirror electron density and the temperature.
Assuming that the galactic mirror plasma is spherical, isothermal and in hydrostatic equilibrium, the temperature is:
\begin{equation}
T = \half \bar{m}v_{\text{rot}}^{2},
\end{equation}
where $\bar{m}$ is the average mass of the mirror ions and the mirror electrons and $\lhaak{v_{\text{rot}}}\approx 220$ km/s is the rotational speed \cite{review}. For a mirror plasma consisting of purely mirror helium, $T \approx 330$ eV. It is convenient to define the electron thermal speed as $v_{0} \equiv \sqrt{2 T/m_{e}}\approx 1.1\times 10^{7}\sqrt{T/\haak{330\text{ eV}}}$ m/s. Boltzmann factors of $\exp\rhaak{-m_{e} v^{2}/\haak{2T}}$ can then be written as $\exp\rhaak{-\haak{v/v_{0}}^{2}}$.

Most light elements except hydrogen that have a large abundance in the universe have a nuclear mass of about twice the proton mass. Assuming that this is also the case for mirror matter, we can relate the mirror electron number density, $n_{e}$, to a good approximation to the total mirror dark matter density, $\rho$, and the contribution of mirror hydrogen to $\rho$ which we denote by $\rho_{H}$:
\begin{equation}\label{rhod}
\rho \approx \sum_{j}n_{j}m_{j}=m_{p}\sum_{j}n_{j}\frac{m_{j}}{m_{p}}\approx 2m_{p}n_{e} -\rho_{H}.
\end{equation}
Here the $n_{j}$ denote the number densities of ions with mass $m_{j}$ and $m_{p}$ is the proton mass. If $\rho_{H}$ can be ignored then $n_{e} \approx 0.16/\text{cm}^{3}$ if $\rho = 0.3$ GeV/cm$^{3}$. Later in this article we'll see that $n_{e}$ can be determined experimentally. The hypothesis $\rho \approx 0.3$ GeV/cm$^3$ and $\rho_{H}\approx 0$ can thus be tested. More generally, upper and lower bounds for $\rho_{H}$ can be set.

The plasma angular frequency for $n_{e} \approx 0.16/\text{cm}^3$ is $\omega_{\text{p}}=\sqrt{n_{e} e^2/\haak{m \varepsilon_{0}}} \approx 23\times 10^{3}$ Hz. The Debye screening length is $\lambda_{\text{D}} = 2^{-1/2} v_{0}/\omega_{\text{p}}\approx 340$ m. The frequency of Coulomb collisions is $\sim\omega_{\text{p}}\log\haak{\Lambda}/\Lambda$ where $\Lambda$ which is proportional to the average number of particles in a Debye sphere: $\Lambda = 4\pi\lambda_{\text{D}}^3 n_{\text{e}}$ \cite{plasrev}. Since $\Lambda\approx 6\times 10^{13}$,  the plasma is nearly collisionless. In Table \ref{tabplas} we have listed the values for the plasma parameters that we'll use in this article.
\begin{table}
\begin{center}
\caption{Properties of the mirror plasma assumed in this article.}\label{tabplas}
\begin{tabular}{|l|l|}\hline
Composition & Mirror helium\\\hline
Temperature & T = 330 eV\\\hline
Mirror electron thermal speed & $v_{0}=1.1\times 10^{7}$ m/s\\\hline
Mirror electron density & $n_{e}=0.16\text{ cm}^{-3}$\\\hline
Plasma angular frequency & $\omega_{\text{p}}=23\times 10^{3}$ Hz\\\hline
Debye screening length & $\lambda_{\text{D}}=340$ m\\\hline
Halo wind speed & $\lhaak{v_{\text{h}}}=220$ km/s\\\hline
\end{tabular}
\end{center}
\end{table}

We can roughly estimate the magnitude of polarization effects as follows.
If we place an ordinary electric charge $q$ in a mirror plasma then, ignoring any distortions due to the motion w.r.t. the plasma, the mirror electric potential will be:
\begin{equation}\label{vdeb}
V\haak{r}= \epsilon\frac{q}{4\pi\varepsilon_{0}\lhaak{r}}\exp\haak{-\frac{\lhaak{r}}{\lambda_{\text{D}}}}.
\end{equation}
Consider a capacitor consisting of two infinite conducting plates. If an electric potential difference is maintained between the two plates, then it follows from a straightforward integration of \eqref{vdeb} that outside the capacitor, close to one of the plates, the induced mirror charge density is:
\begin{equation}\label{rhdeb}
\rho = -\epsilon\varepsilon_{0}\nabla^2 V = -\frac{\epsilon\varepsilon_{0}}{2}\frac{\Delta V}{\lambda_{\text{D}}^{2}}.
\end{equation}
Here $\Delta V$ is the potential of the plate closest to the point where we evaluate $\rho$ relative to the plate furthest from that point and we have assumed that the distance between the two plates is much smaller than $\lambda_{\text{D}}$. This mirror charge density will also exist inside a superconducting cavity placed close to the capacitor and will give rise to electric fields in the interior of the cavity. In case of a cylindrical cavity with a cylindrical conductor at the center, the ordinary electric potential difference between the outer and inner conductor, $\Delta V_{\text{cav}}$ is:
\begin{equation}\label{vcav}
\Delta V_{\text{cav}} = \frac{\epsilon^{2}}{8}\frac{r_{2}^2 - r_{1}^2}{\lambda_{\text{D}}^{2}}\Delta V.
\end{equation}
Here $r_{2}$ and $r_{1}$ are the radii of the outer and inner conductors, respectively. Electric field strengths between the capacitor plates of order $5\times 10^{6}$ V/m can be easily achieved if a material with a high dielectric strength such as e.g.\ polyethylene is used as a dielectric medium. Taking the radius of the cylindrical cavity  $r_{2}\sim 1$ m and the distance between the capacitor plates $\sim 10$ m, then values for $\Delta V_{\text{cav}}$ of about $10^{-15}$ V can be obtained if $\lhaak{\epsilon} \sim 5\times 10^{-9}$ (the value favored by the mirror world explanation of the DAMA results). Static electric fields of this magnitude are difficult to detect. We can, however, let $\Delta V$ oscillate at some angular frequency $\omega$. This will induce an emf inside the cavity. In the next section we will see that the emf is approximately given by \eqref{vcav} for $\omega\lesssim\omega_{p}$ and falls off to zero for $\omega\rightarrow\infty$. The emf has a large internal impedance caused by the small capacitance of the cavity. Suppose that the cylindrical cavity is curved into a toroidal shape with length $l\gg r_{2}$. Such a cavity has a capacitance $C$ of:
\begin{equation}
C=2\pi\varepsilon_{0}\frac{l}{\log\haak{\frac{r_{2}}{r_{1}}}}\approx\frac{1.67\haak{\frac{l}{30\text{ m}}}}{\log{\haak{\frac{r_{2}}{r_{1}}}}}\text{ nF}.
\end{equation} 
To extract the maximum amount of power from this emf one can use a superconducting coil with high $Q$ factor connected between the inner and outer conductors to cancel out the imaginary part of the impedance of the emf. The self inductance of the coil needs to be:
\begin{equation}
L_{\text{coil}} = \frac{1}{\omega^{2}C} = 1.5 \haak{\frac{20\text{ KHz}}{\omega}}^{2}\haak{\frac{30\text{ m}}{l}}\log\haak{\frac{r_{2}}{r_{1}}}\text{ H}.
\end{equation}
The quality factor of such a coil can be about $Q\sim 10^{6}$ at frequencies in the KHz range \cite{coil}. The losses in the superconducting cavity walls are very low and can be ignored; $Q$ can be $\sim 10^{10}$ for resonant cavities operating at GHz frequencies \cite{qcav} and unlike in case of non-superconducting cavities, the $Q$ factor becomes higher at lower frequencies. The power that can be extracted from the emf is thus given by:
\begin{equation}\label{est}
\begin{split}
 P & = \frac{Q C\omega}{4}\haak{\Delta V_{\text{cav}} }^{2}
\approx  \haak{\frac{\epsilon}{5\times 10^{-9}}}^{4}\haak{\frac{340\text{ m}}{\lambda_{\text{D}}}}^{4}\haak{\frac{r_{2}}{\text{m}}}^{4}\times\\ &\haak{\frac{\Delta V}{5\times 10^{7}\text{ V}}}^{2}\haak{\frac{\omega}{20\text{ KHz}}}\haak{\frac{l}{30\text{ m}}}\haak{\frac{Q}{10^{6}}}\frac{1}{\log\haak{\frac{r_{2}}{r_{1}}}}\times\\
& 1.5\times 10^{-29}\text{ W}.
\end{split}
\end{equation}
If a noise temperature of $T_{\text{noise}}\sim 10$ K can be achieved then the polarization effect can be detected after an integration time of $4 k_{\text{b}} T_{\text{noise}} /P\sim 400$ days if $\lhaak{\epsilon}\sim 5\times 10^{-9}$. 

Results from this experiment can also be used to significantly lower the upper limit on the photon mass. Polarization effects from the mirror plasma are similar to the effects of a photon mass of $\lhaak{\epsilon}/\lambda_{\text{D}}$ on length scales much smaller than $\lambda_{\text{D}}$. As explained in Ref.\ \cite{vort}, if the photon acquires a mass due to the Higgs effect then the sharp limits posed by the galactic vector potential are invalid. In that case the sharpest limit on the photon mass is given by the most precise experimental test of Coulomb's law to date \cite{photlim}, which implies an upper limit on the photon mass of $\sim 10^{-14}$ eV. This corresponds to an upper limit on $\lhaak{\epsilon}$ of $2\times 10^{-5}$. Testing the mirror world explanation of the DAMA annual modulation signal would lower the upper limit on the photon mass to $5\times 10^{-9}/\haak{340\text{ m}}\approx 3\times 10^{-18}$ eV.

Equation \eqref{rhdeb} will start to break down when $\omega\sim\omega_{\text{p}}$. To find the polarization effect for general frequencies, we will set up a derivation based on the Vlasov equation. This will also allow for a calculation of the diurnal modulation caused by the anisotropic mirror electron velocity distribution in the Earth's frame.

\section{Vlasov equation}\label{vlasov}
The dynamics of a collisionless plasma is described by the Vlasov equation \cite{plasrev}:
\begin{equation}\label{vl1}
\frac{\partial f}{\partial t} + v \cdot \nabla f -\frac{e}{m_{\text{e}}}\haak{E + v\times B}\cdot
\nabla_{v}f =0.
\end{equation}
Here $f$ is the mirror electron distribution function defined as the density of mirror electrons per unit phase-space volume. This has to be supplemented with the Maxwell equations for the mirror electromagnetic fields. We will ignore the generation of magnetic fields by mirror electron and mirror ion currents. These effects are suppressed by a factor of order $\lhaak{v_{\text{h}}}/c$ relative to the main effect. The effect of the displacement current is suppressed by an additional factor of order (system size) $\times\omega/c$ and will also be ignored. Within these approximations, the inhomogeneous Maxwell equations are:
\begin{equation}\label{mxw}
\begin{split}
&\nabla\cdot E = \frac{e}{\varepsilon_{0}}\rhaak{2 n_{\text{ion}}-\int f d^3 v} + \epsilon\frac{\rho_{\text{ext}}}{\varepsilon_{0}},\\
& \nabla\times B = \epsilon\mu_{0}J_{\text{ext}}.
\end{split}
\end{equation}
Here $n_{\text{ion}}$ is the local mirror helium ion density, $f_{\text{ion}}$ is the velocity distribution of the mirror helium ions normalized to the local density, $\rho_{\text{ext}}$ and $J_{\text{ext}}$ are the ordinary charge and current densities, respectively. They satisfy the equations:
\begin{equation}
\begin{split}
& \rho_{\text{ext}} = \varepsilon_{0}\nabla\cdot E_{\text{ext}},\\
& J_{\text{ext}} =\frac{1}{\mu_{0}}\nabla\times B_{\text{ext}}.
\end{split}
\end{equation}
where $E_{\text{ext}}$ and $B_{\text{ext}}$ are the ordinary 'external' fields that are used to polarize the mirror plasma.

Treating the external fields as a perturbation, the mirror electron distribution function can be written to first order as $f = f_{0} + f_{1}$ where $f_{0}$ is the unperturbed distribution function and $f_{1}$ is the first order correction caused by the external fields. The unperturbed mirror electron distribution function is:
\begin{equation}\label{f0}
f_{0}\haak{v} = n_{\text{e}}\haak{\pi v_{0}^{2}}^{-3/2}\exp\rhaak{-\frac{\haak{v - v_{\text{h}}}^2}{v_{0}^{2}}}.
\end{equation}
Here $v_{\text{h}}$ is the average velocity of the plasma ions and electrons w.r.t.\ Earth. In case of a non-rotating plasma, $\lhaak{v_{\text{h}}}=v_{\text{rot}}\approx 220$ km/s.
From the Vlasov equation \eqref{vl1} and the Maxwell equations \eqref{mxw} we obtain:
\begin{equation}\label{f1}
\begin{split}
\frac{\partial f_{1}}{\partial t} + v\cdot\nabla f_{1} - \frac{e}{m_{\text{e}}}\haak{E + v\times B}\cdot \nabla_{v}f_{0}=0, & \\
\begin{split}
\nabla\cdot E &= \epsilon\nabla\cdot E_{\text{ext}}-\frac{e}{\varepsilon_{0}}\int d^{3} v f_{1},\\
 B &= \epsilon B_{\text{ext}}.
\end{split}&
\end{split}
\end{equation}

In appendix \ref{solvlas} we eliminate $f_{1}$ from \eqref{f1}. If the external fields are proportional to $\sin\haak{\omega t}$ then we find for the mirror charge density:
\begin{equation}\label{rslt}
\begin{split}
\rho\haak{x,t}=&\epsilon\rho_{\text{ext}}\haak{x,t} +\\ & \epsilon\int d^{3}y A\haak{y}\rhaak{G\haak{x-y,\omega,t}-G\haak{x-y,-\omega,t}},
\end{split}
\end{equation}
where $A\haak{y}$ is the coefficient of $\sin\haak{\omega t}$ in the 'effective charge':
\begin{equation}\label{vcrosb}
\rho_{\text{ext}}\haak{y,t}+\varepsilon_{0}\nabla\cdot\haak{v_{\text{h}}\times B_{\text{ext}}\haak{y,t}}\equiv A\haak{y}\sin\haak{\omega t}.
\end{equation}
The Green's function $G\haak{x,\omega,t}$ is defined as:
\begin{equation}
\begin{split}
G\haak{x,\omega,t}\equiv\frac{\exp\haak{i\omega t}}{2i\haak{2\pi}^{3}}&\int d^{3}k\exp\haak{i k\cdot x}\times 
\\ &\rhaak{\frac{2\lambda_{\text{D}}^{2}k^{2}}{Z'\haak{-\frac{\omega}{\sqrt{2}\omega_{\text{p}}\lambda_{\text{D}}\lhaak{k}}-\frac{v_{\text{h}}\cdot k}{v_{0}\lhaak{k}}}}-1}^{-1},
\end{split}
\end{equation}
where $Z'\haak{\zeta}$ is the derivative of the plasma dispersion function \cite{cont} and can be expressed as:
\begin{equation}
Z'\haak{\zeta}=-2 + 2 \sqrt{\pi}\zeta\exp\haak{-\zeta^{2}}\haak{\erfi\haak{\zeta}-i}.
\end{equation}
Here we have defined $\erfi\haak{\zeta}$ as 
\begin{equation}
\erfi\haak{\zeta}\equiv\frac{\erf\haak{i\zeta}}{i}=\frac{2}{\sqrt{\pi}}\int_{0}^{\zeta}dt\exp\haak{t^{2}}.
\end{equation}

We will now apply this to the case of a perturbation caused by a capacitor. Unlike in the case $\omega\ll\omega_{\text{p}}$, we can't ignore finite size effects here. In practice, the size of the capacitor will be much smaller than $\lambda_{\text{D}}$. This means that plasma resonance effects are strongly suppressed. To simplify the calculations we'll take the capacitor plates to be disk shaped. From the estimate of the power of the signal at low frequencies \eqref{est}, we can see that the length of the cavity must be of the order of $\sim 30$ meters. If we take the cavity to be toroidal with its center on the symmetry axis of the capacitor, the mirror charge density is measured at $\sim 5$ meters off the symmetry axis. If we choose the radius of the capacitor plates to be a few tens of meters then this will to a good approximation be the same as the charge density on the symmetry axis. Because we need to reserve some space between the capacitor and the cavity to allow for the necessary shielding and cooling apparatus, the mirror charge density should be evaluated a few meters from the capacitor. Evaluating the mirror charge density on the symmetry axis at a finite distance from the capacitor will thus allow for realistic estimates of what one can expect in experiments.

\begin{figure}
\setlength{\unitlength}{0.07 \textwidth}
\begin{center}
\begin{picture}(5,10)
\put(2.5,7,5){\makebox(0,0){
\begin{picture}(5,5)
\put(2.5,1.5){\makebox(0,0){\includegraphics[width=0.4\textwidth]{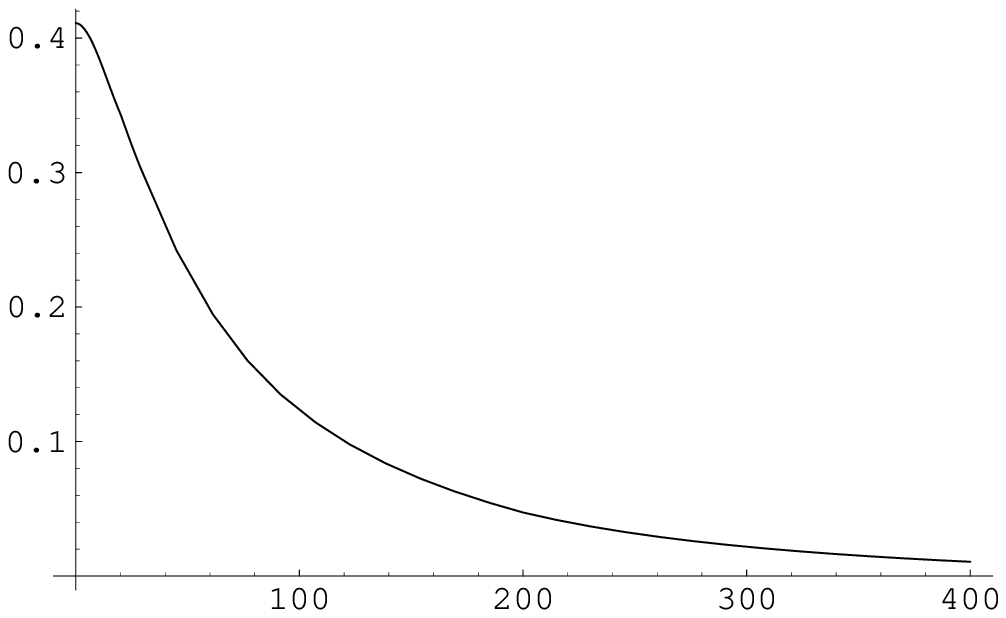}}}
\put(0,3.4){\makebox(0,0){$a_{1}$}}
\put(5.6,0){\makebox(0,0){$\frac{\omega}{\omega_{\text{p}}} $}}
\end{picture}}}

\put(2.5,2.5){\makebox(0,0){
\begin{picture}(5,5)
\put(2.5,1.5){\makebox(0,0){\includegraphics[width=0.4\textwidth]{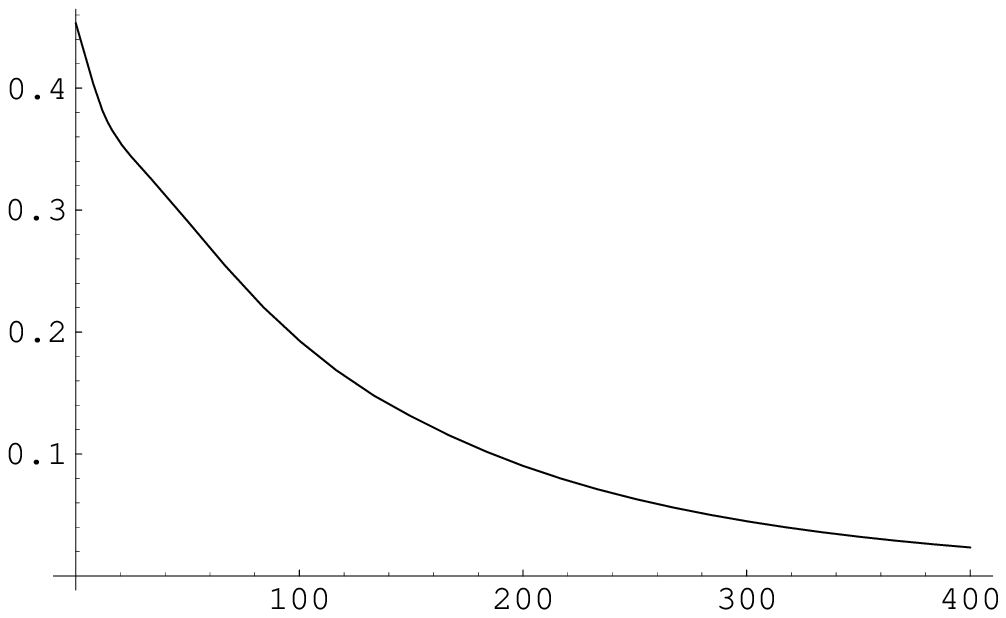}}}
\put(0,3.4){\makebox(0,0){$a_{2}$}}
\put(5.6,0){\makebox(0,0){$\frac{\omega}{\omega_{\text{p}}} $}}
\end{picture}}}
\end{picture}
\caption{The amplitudes $a_{1}$ and $a_{2}$ as a function of the frequency for the case $R=50$ m, $L=10$ m and $\lhaak{d}=4$ m.}\label{amp}
\end{center}
\setlength{\unitlength}{1 pt}
\end{figure}

\begin{figure}
\setlength{\unitlength}{0.07 \textwidth}
\begin{center}
\begin{picture}(5,10)
\put(2.5,7.5){\makebox(0,0){
\begin{picture}(5,5)
\put(2.5,1.5){\makebox(0,0){\includegraphics[width=0.4\textwidth]{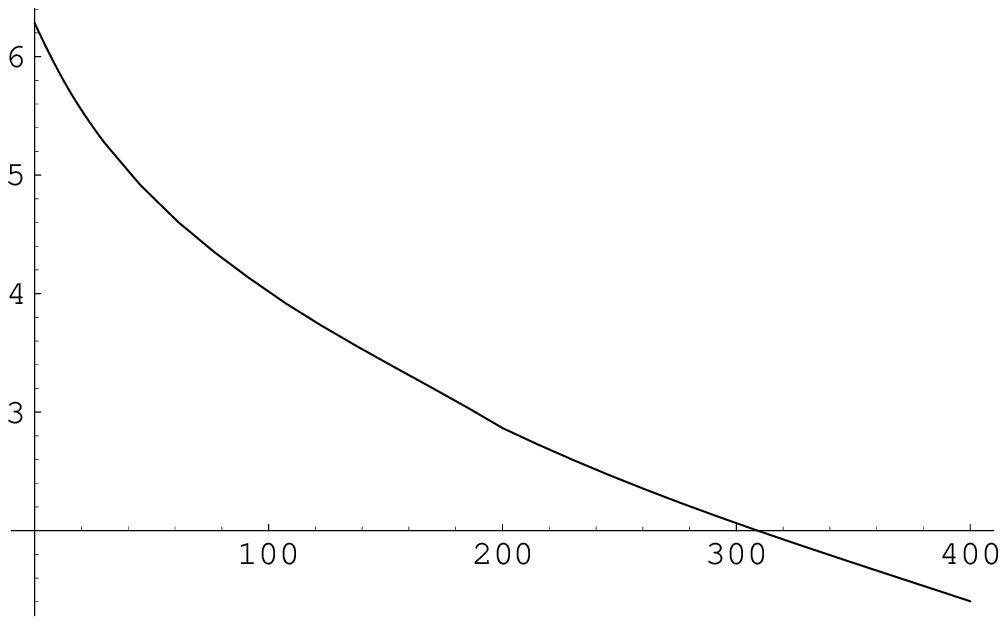}}}
\put(-0.2,3.4){\makebox(0,0){$\phi_{1}$}}
\put(5.6,0.2){\makebox(0,0){$\frac{\omega}{\omega_{\text{p}}} $}}
\end{picture}}}

\put(2.5,2.5){\makebox(0,0){
\begin{picture}(5,5)
\put(2.5,1.5){\makebox(0,0){\includegraphics[width=0.4\textwidth]{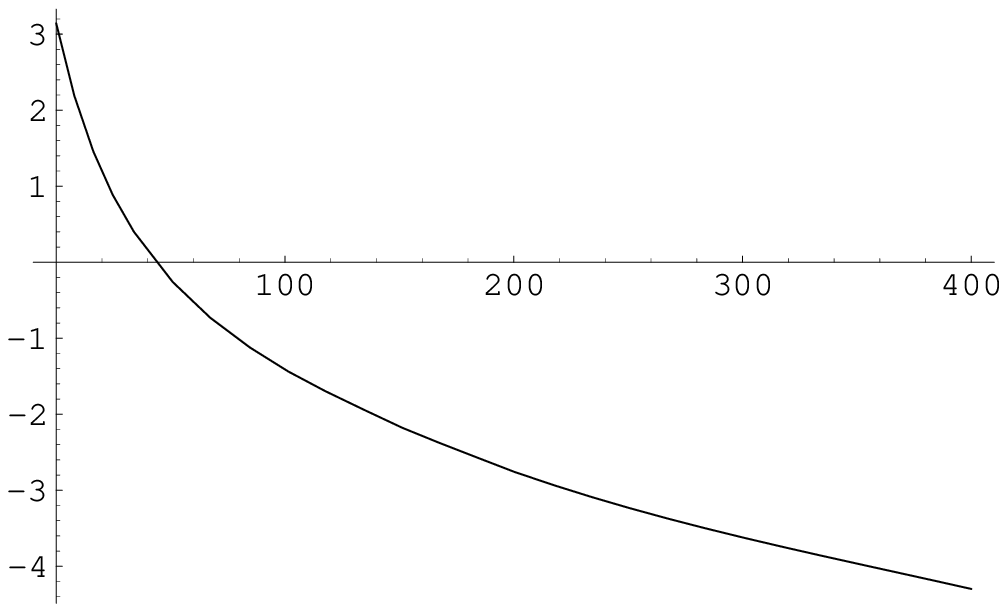}}}
\put(-0.1,3.4){\makebox(0,0){$\phi_{2}$}}
\put(5.6,1.7){\makebox(0,0){$\frac{\omega}{\omega_{\text{p}}} $}}
\end{picture}}}
\end{picture}
\caption{The phase shifts $\phi_{1}$ and $\phi_{2}$ as a function of the frequency for the case $R=50$ m, $L=10$ m and $\lhaak{d}=4$ m. At zero frequency $\phi_{1}=0\mod\haak{2\pi}$ and $\phi_{2}=\pi\mod\haak{2\pi}$.}\label{phs}
\end{center}
\setlength{\unitlength}{1 pt}
\end{figure}

Because $\lhaak{v_{\text{h}}}/v_{0}\approx 1/50$, we can obtain a good approximation for the mirror charge density by expanding it in this parameter to first order. If the potential of the plate closest to the cavity relative to the other plate is $\Delta V\sin\haak{\omega t}$ then we find:
\begin{equation}\label{rhoai}
\rho\haak{d,\omega,t}=-\frac{\epsilon\varepsilon_{0} \Delta V}{\lambda_{\text{D}}^{2}}\rhaak{A_{1}\haak{d,\omega,t} + \frac{\lhaak{v_{\text{h}}}}{v_{0}}\cos\haak{\theta}A_{2}\haak{d,\omega,t}}.
\end{equation}
Here $d$ denotes the position where $\rho$ is evaluated relative to the center of the closest capacitor plate and $\theta$ is the angle between $v_{\text{h}}$ and $d$. The dimensionless functions $A_{i}$ are given by:
\begin{equation}
\begin{split}
A_{i}\haak{d,\omega,t}=\frac{R}{2\pi L}\Im\int_{0}^{\infty}du\int_{0}^{\pi/2}d\beta u J_{1}\haak{\frac{R}{\lambda_{\text{D}}}u\sin\haak{\beta}}&\times \\ \rhaak{\exp\haak{i u\cos\haak{\beta}\frac{L+\lhaak{d}}{\lambda_{\text{D}}}}-\exp\haak{i u\cos\haak{\beta}\frac{\lhaak{d}}{\lambda_{\text{D}}}}}&\times \\ \rhaak{G_{i}\haak{u,\frac{\omega}{\omega_{\text{p}}}}\exp\haak{i\omega t}-G_{i}\haak{u,-\frac{\omega}{\omega_{\text{p}}}}\exp\haak{-i\omega t}}.&
\end{split}
\end{equation}
Here $L$ is the distance between the plates, $R$ is the radius of the plates and the $G_{i}$ are given by:
\begin{gather}
\begin{split}
& G_{1}\haak{u,f}=\\ &\frac{-i\sqrt{2\pi}f+2u\exp\haak{\frac{f^{2}}{2u^{2}}}-\sqrt{2\pi}f\erfi\haak{\frac{f}{\sqrt{2}u}}}{i\sqrt{2\pi}f-2\haak{u+u^{3}}\exp\haak{\frac{f^{2}}{2u^{2}}}+\sqrt{2\pi}f\erfi\haak{\frac{f}{\sqrt{2}u}}},
\end{split}\\
\begin{split}
& G_{2}\haak{u,f}=-4u^{2}\exp\haak{\frac{f^{2}}{2u^{2}}}\times\\ &\frac{i\sqrt{\pi}f^{2}-\sqrt{2}fu\exp\haak{\frac{f^{2}}{2u^{2}}}-i\sqrt{\pi}u^{2}+\sqrt{\pi}\haak{f^{2}-u^{2}}\erfi\haak{\frac{f}{\sqrt{2}u}}}{\rhaak{i\sqrt{2\pi}f-2\haak{u+u^{3}}\exp\haak{\frac{f^{2}}{2u^{2}}}+\sqrt{2\pi}f\erfi\haak{\frac{f}{\sqrt{2}u}}}^{2}}.
\end{split}
\end{gather}
We have numerically calculated the functions $A_{i}$ as a function of $\omega$ for the case $R=50$ m, $L=10$ m and $\lhaak{d}=4$ m. We represent these in Figs.\ \ref{amp} and \ref{phs} by their amplitudes $a_{i}$ and phase shifts $\phi_{i}$ relative to the external field:
\begin{equation}\label{ait}
A_{i}\haak{d,\omega,t}\equiv a_{i}\haak{d,\omega}\sin\rhaak{\omega t +\phi_{i}\haak{\omega}}.
\end{equation}

The $\cos\haak{\theta}$ factor multiplying $A_{2}$ in \eqref{rhoai} is a periodic function with a period of one sidereal day which is determined by the Earth's motion relative to the galactic frame and the orientation of the capacitor. If the capacitor plates are placed parallel to the Earth's surface and the cavity is placed below the bottom plate then we can write $\cos\haak{\theta}$ as the inner product of the unit vector pointing in the direction of $-v_{\text{h}}$ and the unit vector pointing from the Earth's center toward the detector:
\begin{equation}\label{thetat}
\cos\haak{\theta}=\cos\haak{\delta}\cos\haak{\alpha}+\sin\haak{\delta}\sin\haak{\alpha}\cos\rhaak{\Omega\haak{t-t_{0}}},
\end{equation}
where $\delta\approx 43^{\circ}$ is the angle of $-v_{\text{h}}$ with the vector parallel to the Earth's axis, $\alpha$ is the angle between the Earth's axis and the position of the detector relative to the center of the Earth and can be expressed as $90^{\circ}$ minus the geographical latitude, $\Omega=2\pi$ (sidereal day)$^{-1}$ and $t_{0}\approx 21$ hours local sidereal time so that $\Omega\haak{t-t_{0}}$ is the angle between the vector $-v_{\text{h}}$ projected onto the Earth's equatorial plane and the position of the detector relative to the Earth's center projected onto this plane. In appendix \ref{trig} we show how these values for $\delta$ and $t_{0}$ are obtained from the orbital parameters of the motion of the Earth around the Sun and the Sun around the Milky Way.

If we insert \eqref{ait} and \eqref{thetat} in \eqref{rhoai} and expand to first order in $\lhaak{v_{\text{h}}}/v_{0}$ we obtain:
\begin{equation}\label{rhot}
\begin{split}
\rho\haak{d,\omega,t}=&-\frac{\epsilon\varepsilon_{0} \Delta V}{\lambda_{\text{D}}^{2}} 
 \ahaakl{b_{1}\sin\haak{\omega t+\varphi_{1}}+\vphantom{b_{2}\rhaak{\sin\haak{\haak{\omega-\Omega} t+\varphi_{2}}+\sin\haak{\haak{\omega+\Omega}t+\varphi_{3}}}}}\\
&\ahaakr{\vphantom{b_{1}\sin\haak{\omega t+\varphi_{1}}+} b_{2}\rhaak{\sin\haak{\haak{\omega-\Omega} t+\varphi_{2}}+\sin\haak{\haak{\omega+\Omega}t+\varphi_{3}}}},
\end{split}
\end{equation}
where
\begin{align}
 b_{1}=& a_{1} + a_{2}\frac{\lhaak{v_{\text{h}}}}{v_{0}}\cos\haak{\delta}\cos\haak{\alpha}\cos\haak{\phi_{2}-\phi_{1}},\\
 b_{2}=&\frac{a_{2}}{2}\frac{\lhaak{v_{\text{h}}}}{v_{0}}\sin\haak{\delta}\sin\haak{\alpha},
\end{align}
and
\begin{align}
\varphi_{1}=&\phi_{1}+\frac{a_{2}}{a_{1}}\frac{\lhaak{v_{\text{h}}}}{v_{0}}\cos\haak{\delta}\cos\haak{\alpha}\sin\haak{\phi_{2}-\phi_{1}},\\
\varphi_{2}=&\phi_{2}+\Omega t_{0},\\
\varphi_{3}=&\phi_{2}-\Omega t_{0}.
\end{align}
Comparing \eqref{rhot} with \eqref{rhdeb}, we see that the power of the signal at frequency $\omega$ is $4 b_{1}^2 P\approx 4 a_{1}^{2} P$  where $P$ is given by \eqref{est}\footnote{Here we ignore the fact that the mirror charge density inside the cavity is not constant and depends on the distance to the capacitor.}. The signal at the frequencies $\omega\pm \Omega$ is about $10^{4}$ times weaker.

From Eq.\ \ref{vcrosb} we see that strong magnetic fields of order $\lhaak{B}\sim 10$ T will have similar effects as electric fields of order $\lhaak{v_{\text{h}}} \lhaak{B}\sim 2\times 10^{6}$ V/m. This is similar in magnitude as in the case of the capacitor we have considered; experiments using large coils can thus yield similar results as experiments using large capacitors. In the case of external magnetic fields the leading polarization effect will depend on the time averaged magnitude of $v_{\text{h}}\times B$. By comparing measurements using electric and magnetic fields one can thus measure both the direction and the magnitude of $v_{\text{h}}$. The order $\lhaak{v_{\text{h}}}/v_{0}$ correction of the main polarization effect in external electric fields can be measured by performing experiments at different geographic latitudes. This thus allows one to determine $v_{0}$. Once this parameter is fixed, the frequency profile of the polarization effect depends, up to an overall coefficient proportional to $\epsilon^{2}$, only on $\omega_{\text{p}}$ because $\lambda_{\text{D}}=2^{-1/2}v_{0}/\omega_{\text{p}}$. A fit to the frequency profile is thus sufficient to determine $\omega_{\text{p}}$, $\lambda_{\text{D}}$ and $\lhaak{\epsilon}$.

If $\lambda_{D}$ is much larger than the size of the capacitor or the coil, then it will be difficult to determine $\omega_{\text{p}}$ and $\lambda_{\text{D}}$. In that case the polarization effect becomes almost invariant under the rescaling: $\lambda_{\text{D}}\rightarrow S \lambda_{\text{D}}$, $\omega_{\text{p}}\rightarrow \omega_{\text{p}}/S$, and $\epsilon\rightarrow S\epsilon$. If $\omega_{\text{p}}$ can be measured then that fixes the mirror electron density and, as explained in Section \ref{secpol}, this allows one to infer the mirror dark matter density (see Eq.\ \ref{rhod}). Effects due to mirror matter can be easily distinguished from effects due to a finite photon mass because in the latter case the frequency profile is flat and the phase shift is zero.

\section{Discussion}
We have shown that in strong electromagnetic fields mirror dark matter can become polarized and give rise to electric fields in cryogenic cavities. This effect exists for frequencies ranging from zero up to a few MHz. Using this effect all the parameters of the mirror dark matter plasma can in principle be measured. Besides detecting or constraining mirror dark matter, this experiment can also significantly improve the upper limit on the photon mass.

If the DAMA annual modulation signal is due to mirror dark matter then the expected signal strength, given by Eq.\ \ref{est}, is very small and a large cavity is therefore needed. The main difficulty in such a large scale experiment will be to keep the cavity at cryogenic temperatures.
The required cryogenics facilities would have to be similar to those for the magnets of the ATLAS experiment at the LHC \cite{atlas}.

There are however scenarios which can yield a positive result in less sensitive, smaller scale experiments. From Eq.\ \ref{est} we see that the signal strength is proportional to $\epsilon^{4}$, so mirror matter scenarios with a larger value for $\lhaak{\epsilon}$ than the favored value of $\lhaak{\epsilon}\sim 5\times 10^{-9}$ would yield much stronger signals. The BBN bound on $\lhaak{\epsilon}$ is $3\times 10^{-8}$ \cite{nucleo}, but increasing $\lhaak{\epsilon}$ to this value would require lower abundances of elements heavier than helium by a factor of about 100 to avoid conflict with the DAMA results. This is problematic because we need to assume a high rate of supernova explosions to keep the mirror plasma hot enough to prevent it from collapsing into a disk \cite{ref}.

It may be possible to evade these difficulties in broken mirror matter models \cite{brok}. Here each mirror particle is heavier than its ordinary partner by some constant factor. This causes the rate of cooling of the halo to be slower than in the unbroken case; the time needed for the halo to collapse is $\sim 10^{8}$ years times the mass ratio squared \cite{brok2}. If a photon-mirror photon kinetic mixing also exists in this case then this type of dark matter can be detected in electromagnetic field penetration experiments. The signal can be much stronger than in the unbroken case if conflicts with dark matter direct detection experiments can be avoided. This is possible if the mirror elements are not heavier than about 15 atomic mass units. As explained in Ref.\ \cite{brok3}, the mirror neutron can be unstable, even if bound in nuclei. As a consequence, mirror hydrogen would then be the only stable mirror element. This means that if the mirror masses are 10 to 15 times larger then one can have both a spherical halo and no signal in dark matter direct detection experiments. The DAMA signal would then have to be due to a different component of the dark matter or not due to dark matter at all. Because mirror electrons are heavier in this case, the BBN bound on $\lhaak{\epsilon}$ can be relaxed a bit \cite{footcom}. Also there is no positronium-mirror positronium oscillation in this case.

But whatever the theoretical motivations are for doing this experiment, the fact remains that the strongest null result to date comes from the most precise direct test of Coulomb's law \cite{photlim}, which corresponds to an effective upper limit on $\lhaak{\epsilon}$ of $2\times 10^{-5}$. This means that unknown physics could yield signal strengths of up to $\sim 10^{14}$ times the estimate given in Eq.\ \ref{est}!

\section{Acknowledgments}
I thank Robert Foot, Alex Ignatiev, Zurab Silagadze and the referee for their comments on this article.

\appendix
\section{Solving the Vlasov equation}\label{solvlas}
In this section we'll solve the Vlasov equation to first order in perturbation theory. The methods used here can be found in books on plasma physics, such as e.g.\ Refs.\ \cite{plasrev,cont}. In section \ref{vlasov} we obtained the first order result \eqref{f1}:
\begin{equation}\label{f1ap}
\begin{split}
\frac{\partial f_{1}}{\partial t} + v\cdot\nabla f_{1} - \frac{e}{m_{\text{e}}}\haak{E + v\times B}\cdot \nabla_{v}f_{0}=0, & \\
\begin{split}
\nabla\cdot E &= \epsilon\nabla\cdot E_{\text{ext}}-\frac{e}{\varepsilon_{0}}\int d^{3} v f_{1},\\
 B &= \epsilon B_{\text{ext}},
\end{split}&
\end{split}
\end{equation}
where $f_{0}$ is the unperturbed mirror electron distribution given by Eq.\ \ref{f0} and $f_{1}$ is the first order correction to the mirror electron distribution.
Attempting to solve these equations by performing a Fourier transformation w.r.t.\ the space and time coordinates leads to singularities. This is caused by the absence of dissipative terms in the Vlasov equations. To deal with this problem one must treat the Vlasov equations as an initial value problem. This is most conveniently done by performing a Fourier transformation w.r.t.\ space and a Laplace transformation w.r.t.\ time. In the following we shall use the notations:
\begin{equation}
\begin{split}
\widehat{F}\haak{k,t}\equiv &\int d^3 x F\haak{x,t}\exp\haak{-i k\cdot x},\\
\overline{F}\haak{k,p}\equiv &\int_{0}^{\infty}dt \widehat{F}\haak{k,t}\exp\haak{- p t}.
\end{split}
\end{equation}

We now assume that at $t=0$, $f_{1}$ and the external fields are zero.
Applying a Fourier-Laplace transformation to \eqref{f1ap} gives:
\begin{equation}\label{f1fl}
\begin{split}
p \overline{f_{1}} + i v\cdot k \overline{f_{1}} - \frac{e}{m_{\text{e}}}\haak{\overline{E} + v\times \overline{B}}\cdot \nabla_{v}f_{0}=0,&\\
\begin{split}
i k\cdot \overline{E} &= \epsilon i k\cdot \overline{E}_{\text{ext}}-\frac{e}{\varepsilon_{0}}
\int d^3 v \overline{f_{1}},\\
\overline{B} &= \epsilon\overline{B}_{\text{ext}}.
\end{split}&
\end{split}
\end{equation}
Eliminating $\overline{f_{1}}$ from these equations yields the mirror charge density in terms of the external fields:
\begin{equation}\label{dive}
\overline{\rho}=\epsilon \overline{\rho}_{\text{ext}}+\epsilon\frac{1}{\frac{2\lambda_{\text{D}}^{2}k^{2}}{I\haak{k,p}}-1}\overline{\rho}_{\text{ext}}^{\text{eff}}.
\end{equation}
Here we have defined $I\haak{k,p}$ as:
\begin{equation}\label{ikp}
I\haak{k,p}\equiv\frac{1}{\sqrt{\pi}}\int_{-\infty}^{\infty}dx\frac{\exp\haak{-x^2}}{\haak{x-\frac{ip}{v_{0}\lhaak{k}}+\frac{v_{\text{h}}\cdot k}{v_{0}\lhaak{k}}}^{2}},
\end{equation}
and $\rho_{\text{ext}}^{\text{eff}}$ is defined as
\begin{equation}
\rho_{\text{ext}}^{\text{eff}}\equiv \varepsilon_{0}\nabla\cdot \rhaak{v_{\text{h}}\times B_{\text{ext}} +E_{\text{ext}}}.
\end{equation}

\begin{figure}
\begin{center}
\begin{picture}(200,200)
\put(100,100){\makebox(0,0){\includegraphics[width=200 pt]{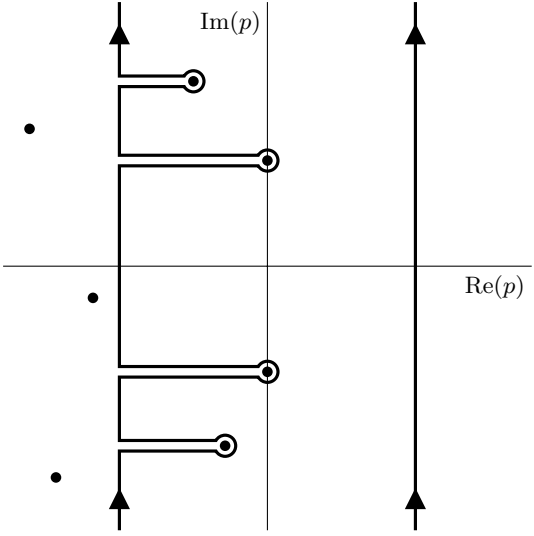}}}
\put(186,92){\makebox(0,0){$\Re\haak{p}$}}
\put(86,192){\makebox(0,0){$\Im\haak{p}$}}
\end{picture}
\caption{Deforming the Bromwich contour. The contour on the right is the Bromwich contour $C$ in Eq.\ \ref{invlap}, the dots are the poles and the deformed contour is shown on the left.}\label{brom}
\end{center}
\end{figure}

We now assume that $\rho_{\text{ext}}^{\text{eff}}$ is of the form:
\begin{equation}
\rho_{\text{ext}}^{\text{eff}}\haak{x,t}=A\haak{x}\sin\haak{\omega t}.
\end{equation}
This means that $\overline{\rho}_{\text{ext}}^{\text{eff}}$ is given by:
\begin{equation}
\overline{\rho}_{\text{ext}}^{\text{eff}}\haak{k,p}=\frac{\widehat{A}\haak{k}}{2i}\rhaak{\frac{1}{p-i\omega}-\frac{1}{p+i\omega}}.
\end{equation}
Inserting this in \eqref{dive} and taking the inverse Laplace transform gives:
\begin{equation}\label{invlap}
\widehat{\rho}=\epsilon \widehat{\rho}_{\text{ext}} + \frac{\epsilon}{2\pi i}\frac{\widehat{A}\haak{k}}{2i}\int_{C}dp\frac{\exp\haak{p t}}{\frac{2\lambda_{\text{D}}^{2}k^{2}}{I\haak{k,p}}-1}\rhaak{\frac{1}{p-i\omega}-\frac{1}{p+i\omega}},
\end{equation}
where $C$ is a Bromwich contour with $\Re\haak{p}>0$. We want to find $\rho$ at large times when  the transient effects, caused by the switching on of the external fields, have vanished. The standard technique is to deform the Bromwich contour by moving it to the left of the imaginary axis but such that the poles are still encircled to the right, see Fig.\ \ref{brom}. Then, for $t\rightarrow\infty$, only the contribution from the poles at $p = \pm i\omega$ will survive. Because the integral defining $I\haak{k,p}$ in Eq.\ \ref{ikp} is singular for $\Re\haak{p} = 0$ we must analytically continue it from $\Re\haak{p}>0$ to the region $\Re\haak{p}\le 0$. This can be done by deforming the integration contour in Eq.\ \ref{ikp} by letting it pass below the pole at $x=\frac{ip}{v_{0}\lhaak{k}}-\frac{v_{\text{h}}\cdot k}{v_{0}\lhaak{k}}$.
Denoting the analytical continuation of $I\haak{k,p}$ by $I^{*}\haak{k,p}$ we can then write:
\begin{equation}\label{rhofour}
\widehat{\rho}=\epsilon \widehat{\rho}_{\text{ext}} + \epsilon\frac{\widehat{A}\haak{k}}{2i}\rhaak{\frac{\exp\haak{i\omega t}}{\frac{2\lambda_{\text{D}}^{2}k^{2}}{I^{*}\haak{k,i\omega}}-1}-\frac{\exp\haak{-i\omega t}}{\frac{2\lambda_{\text{D}}^{2}k^{2}}{I^{*}\haak{k,-i\omega}}-1}}.
\end{equation}
An expression for $I^{*}\haak{k,p}$ in terms of error functions can be found as follows. It is convenient to introduce the plasma dispersion function $Z\haak{\zeta}$ \cite{cont} which is defined as:
\begin{equation}
Z\haak{\zeta}\equiv\frac{1}{\sqrt{\pi}}\int_{C^{-}}dz \frac{\exp\haak{-z^{2}}}{z-\zeta},
\end{equation}
where $C^{-}$ is a contour from $-\infty$ to $+\infty$ that passes below the pole at $z=\zeta$. The function $I^{*}\haak{k,p}$ can then be expressed in terms of the derivative of the plasma dispersion function as
\begin{equation}
I^{*}\haak{k,p}= Z'\haak{\frac{ip}{v_{0}\lhaak{k}}-\frac{v_{\text{h}}\cdot k}{v_{0}\lhaak{k}}}.
\end{equation}
Applying a partial integration to
\begin{equation}
Z'\haak{\zeta}=\frac{1}{\sqrt{\pi}}\int_{C^{-}}dz \frac{\exp\haak{-z^{2}}}{\haak{z-\zeta}^2},
\end{equation}
leads to the differential equation:
\begin{equation}
Z'\haak{\zeta}+2\zeta Z\haak{\zeta} + 2=0.
\end{equation}
Solving this equation and using that $Z\haak{0}=i\sqrt{\pi}$ gives:
\begin{equation}
Z\haak{\zeta}=\sqrt{\pi}\exp\haak{-\zeta^{2}}\rhaak{i-\erfi\haak{\zeta}},
\end{equation}
where we have defined:
\begin{equation}
\erfi\haak{\zeta}\equiv\frac{\erf\haak{i\zeta}}{i}=\frac{2}{\sqrt{\pi}}\int_{0}^{\zeta}\exp\haak{t^{2}} dt.
\end{equation}
We thus see that $Z'\haak{\zeta}$ is given by:
\begin{equation}
Z'\haak{\zeta}=-2 + 2 \sqrt{\pi}\zeta\exp\haak{-\zeta^{2}}\haak{\erfi\haak{\zeta}-i}.
\end{equation}
Finally, applying an inverse Fourier transformation to \eqref{rhofour} gives the result:
\begin{equation}
\begin{split}
\rho\haak{x,t}=&\epsilon\rho_{\text{ext}}\haak{x,t} +\\ & \epsilon\int d^{3}y A\haak{y}\rhaak{G\haak{x-y,\omega,t}-G\haak{x-y,-\omega,t}},
\end{split}
\end{equation}
where
\begin{equation}
\begin{split}
G\haak{x,\omega,t}\equiv\frac{\exp\haak{i\omega t}}{2i\haak{2\pi}^{3}}&\int d^{3}k\exp\haak{i k\cdot x}\times\\ &\rhaak{\frac{2\lambda_{\text{D}}^{2}k^{2}}{Z'\haak{-\frac{\omega}{\sqrt{2}\omega_{\text{p}}\lambda_{\text{D}}\lhaak{k}}-\frac{v_{\text{h}}\cdot k}{v_{0}\lhaak{k}}}}-1}^{-1}.
\end{split}
\end{equation}

\section{Time dependence of {\normalsize $\cos\haak{\theta}$}}\label{trig}
The values for $\delta$ and $t_{0}$ in Eq.\ \ref{thetat} can be obtained from the angle of $60^{\circ}$ the ecliptic plane tilts relative to the normal of the galactic equatorial plane in the direction of motion of the Sun around the Milky Way, the date of 2 July when the component of the Earth's velocity in the direction of the motion of the Sun around the Milky Way is maximal (at that time the Earth moves toward the galactic South Pole) \cite{bin} and the fact that the Earth's axis is tilted by $23.5^{\circ}$ relative to the normal of the ecliptic plane. Expressing $\cos\haak{\delta}$ as the inner product of the unit vectors pointing in the directions of $-v_{\text{h}}$ and the Earth's axis gives:
\begin{equation}\label{cosdelta}
\cos\haak{\delta}=\cos\haak{30^{\circ}}\cos\haak{23.5^{\circ}} + \sin\haak{30^{\circ}}\sin\haak{23.5^{\circ}}\cos\haak{\gamma}.
\end{equation}
Here $\gamma$ is the angle between $-v_{\text{h}}$ projected onto the ecliptic plane and the Earth's axis projected onto this plane. Approximating Earth's orbit as a circle, and using that on 21 June the Earth's axis points toward the Sun while on 2nd June $-v_{\text{h}}$ projected onto the ecliptic is parallel to Earth's orbit, we find that
\begin{equation}
\gamma\approx 90^{\circ}+\frac{19 \text{ days}}{1 \text{ year}} 360^{\circ}\approx 108.7^{\circ}.
\end{equation}
Inserting this in \eqref{cosdelta} gives $\delta\approx 43^{\circ}$.

To compute $t_{0}$, we note that around 21 March when the Earth's axis is orthogonal to the unit vector $s$ pointing from the Earth to the Sun, 0 hour local sidereal time corresponds to the moment when $s$ projected onto the Earth's equatorial plane and the position of the detector projected onto that plane are parallel. The angle between $s$ projected onto the Earth's equatorial plane and $-v_{\text{h}}$ projected onto this plane expressed in hours thus gives $t_{0}$. This angle follows from the inner product of $s$ with $-v_{\text{h}}$ by using the fact that $s$ lies in the intersection of the ecliptic plane and the Earth's equatorial plane:
\begin{equation}
-v_{\text{h}}\cdot s= \cos\haak{60^{\circ}}\cos\haak{\gamma-90^{\circ}}=\sin\haak{\delta}\cos\haak{t_{0}}.
\end{equation}
This, combined with the fact that $t_{0}$ must be in the interval ranging from $270^{\circ}$ and $360^{\circ}$, gives $t_{0}\approx 314^{\circ}\approx 21$ hours.

\end{document}